# High-coherence superconducting qubits made using industry-standard, advanced semiconductor manufacturing

J. Van Damme[1,2], S. Massar[1], R. Acharya[1], Ts. Ivanov[1], D. Perez Lozano[1], Y. Canvel[1], M. Demarets[1,2], D. Vangoidsenhoven[1], Y. Hermans[1], J.G. Lai[1], A. M. Vadiraj[1], M. Mongillo[1], D. Wan[1], J. De Boeck[1,2], A. Potočnik[1*], K. De Greve[1,2]

[1]Imec, Kapeldreef 75, Leuven 3000, Belgium
[2]Department of Electrical Engineering (ESAT), KU Leuven, Leuven 3000, Belgium

The development of superconducting qubit technology has shown great potential for the construction of practical quantum computers. As the complexity of quantum processors continues to grow, the need for stringent fabrication tolerances becomes increasingly critical. Utilizing advanced industrial fabrication processes could facilitate the necessary level of fabrication control to support the continued scaling of quantum processors. However, these industrial processes are currently not optimized to produce high coherence devices, nor are they a priori compatible with the commonly used approaches to make superconducting qubits. In this work, we demonstrate for the first time superconducting transmon qubits manufactured in a 300 mm CMOS pilot line, using industrial fabrication methods, with resulting relaxation and coherence times already exceeding 100 microseconds. We show across-wafer, large-scale statistics studies of coherence, yield, variability, and aging that confirm the validity of our approach. The presented industry-scale fabrication process, using exclusively optical lithography and reactive ion etching, shows performance and yield similar to the conventional laboratory-style techniques utilizing metal lift-off, angled evaporation, and electron-beam writing. Moreover, it offers potential for further upscaling by including three-dimensional integration and additional process optimization using advanced metrology and judicious choice of processing parameters and splits. This result marks the advent of more reliable, large-scale, truly CMOS-compatible fabrication of superconducting quantum computing processors.

## I. Introduction

In the pursuit of quantum computational advantage, and eventually, fault-tolerant, error-corrected quantum hardware, a need for more and better physical qubits with high-fidelity control is apparent. Advances in error correcting codes [1,2] and quantum gate fidelities can reduce the required number of physical qubits. Additionally, increased stability and uniformity of the qubits would reduce the significant control and tuning overhead [3]. However, for practical applications, the number of physical qubits on a quantum computer will most likely still scale beyond a million [4].

Superconducting circuit implementations of quantum bits leverage the scalable nature of solid-state fabrication and have shown tremendous progress in terms of qubit coherence times [4,5] and gate fidelities [6,7]. State-of-the-art demonstrations include error correction [8–12], processors with hundreds of interconnected qubits and initial claims of quantum supremacy and utility [13–15]. These demonstrations are all done with architectures utilizing transmon style qubits [16] with Al/AlO$_x$/Al Josephson junctions (JJ) [17–19], which are consistently fabricated using angled shadow evaporation and metal lift-off. The advantage of this fabrication technique lies in the possibility for in situ fabrication of the JJ and minimal etch damage, achieving high coherence qubits, up to hundreds of microseconds [20–23].

With the scaling requirements of future quantum processors in mind, industry scale fabrication of high coherence qubits utilizing all-optical lithography and exclusively reactive ion etch on 300 mm diameter wafers could be the preferred way forward, similar to the state-of-the-art in advanced CMOS fabrication, and in line with recent developments in silicon quantum dot qubit fabrication [24]. This is especially true in view of the enormous sophistication and process control present in modern industrial semiconductor tooling sets, as well as the knowledge on advanced three-dimensional integration techniques being developed for 300 mm diameter wafers in the pursuit of Moore's law [25].

In this work we demonstrate for the first time superconducting transmon qubits fabricated on 300 mm silicon wafers at the foundry-standard cleanroom of Imec, using industry-standard methods that leverage earlier learnings [26] and demonstrate the real advantages of advanced CMOS fabrication methods in terms of processing splits, process parameter control and advanced metrology to identify critical process steps. We validate our approach with extensive, across-wafer analysis and benchmarking, where we characterized 400 qubits and 12,840 JJ test-structures and report on excellent qubit yield, qubit coherence times, qubit frequency variability, and aging statistics. Our initial result showcases a leap forward in the potential fabrication volume and yield of high-coherence superconducting qubits, which, together with three-dimensional integration developments [27–29], could meet the stringent fabrication demands of a future million-qubit processor.

## II. Design and fabrication.

In an industry-standard 300 mm process, an optical mask delineates the pattern of a single die (24 mm × 28 mm), which is replicated 75 times to constitute the complete 300 mm diameter wafer, illustrated in Figure 1a. Each individual die encompasses twenty sub-dies in our mask, featuring distinct designs of qubits, resonators, and JJ test arrays, as shown in Figure 1b. To assess the quality of the qubits and the variability associated with the fabrication process, parameters such as JJ normal resistance ($R_n$), qubit transition frequencies ($f_{qb}$), relaxation times ($T_1$), and Hahn-echo coherence times ($T_2^e$) are tracked across the wafer. A sub-die design (D1) with five transmon qubits of four different capacitor geometries is selected for qubit coherence and energy relaxation time analysis. Another sub-die design (D2) containing ten qubits with identical capacitors and different JJ area's is selected to monitor the qubit frequency variability. More information on the device designs can be found in Supplementary Table I.

The Josephson junction, a crucial element of a superconducting qubit, is fabricated using a 300 mm technology compatible overlap process [26,30,31]. An example Al/AlO$_x$/Al overlap JJ is visualized with a scanning electron microscopy image in Figure 1c, and a cross-section of the JJ is detailed in a transmission electron microscopy image, shown in Figure 1d.

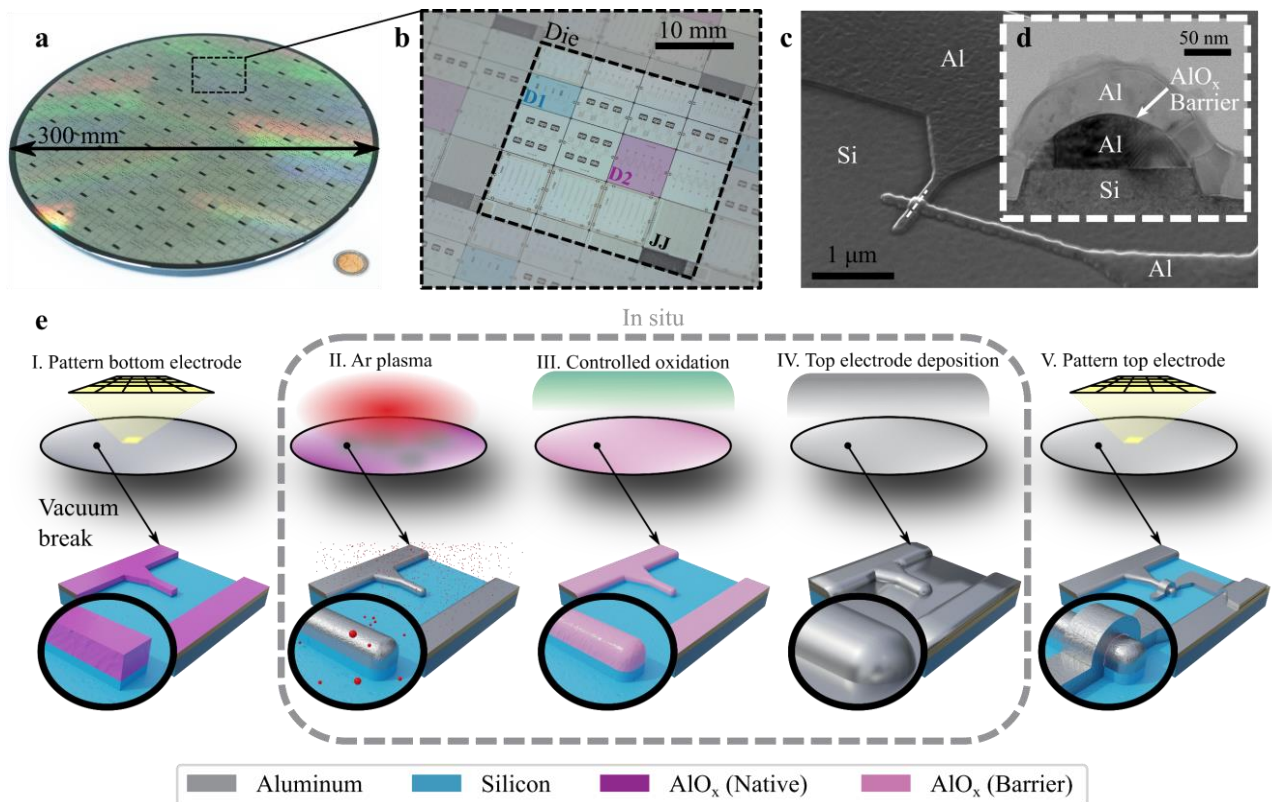

**FIG. 1 | Overlap Josephson junction qubit fabrication. a**, Photograph of the 300 mm wafer. **b**, Photograph of one die with highlights of sub-die designs D1 and D2. **c**, Scanning electron microscopy image of an overlap Josephson junction. **d**, Transmission electron microscopy image of a cross section of the junction (dashed line in **c**). **e**, Schematic representation of the key fabrication steps of overlap Josephson junctions.

The overlap JJ fabrication process of this work is an extension of the methodology described in our prior research [26,32], with the addition of all-optical lithography and the native incorporation of advanced metrology structures in the mask design for process control and monitoring – resulting in improved and more consistent behavior of the devices. Figure 1e provides a schematic representation of the critical steps in the process. In the initial stage (step I in Figure 1e), the bottom electrode (BE) of the JJ is patterned alongside other circuit components such as resonators and ground planes in the first design layer. A 70 nm Al film is sputtered at room temperature (RT) on a hydrofluoric-acid-cleaned, high-resistivity ($R_s \geq 3\,\mathrm{k\Omega cm}$) Si substrate. The first design layer is patterned via optical immersion lithography (193 nm). After exposure, the pattern is transferred to the Al film by subtractive Cl-based dry etch, followed by a diluted sulfuric acid-peroxide wet clean and a de-ionized-water rinse. In the second stage (steps II-IV in Figure 1e), the barrier formation process involves the complete removal of the native Al and Si oxide through an argon milling procedure, succeeded by a regulated dynamic oxidation. This AlO$_x$ barrier is then overlaid in situ with a 50 nm, RT sputtered, Al film. In the final stage (step V in Figure 1e), the top electrode is patterned in the second design layer using an analogous process to the BE patterning with optical immersion lithography and dry etch. Upon completion of fabrication, the wafer is coated with a protective resist layer prior to dicing. This resist is subsequently removed from the sub-dies using acetone, isopropanol, and a brief 2-min oxygen ashing process. Finally, the sub-dies undergo a post-etch residue removal treatment (EKC) before they are wire-bonded into a measurement package.

### III. Qubit coherence.

The described fabrication process is benchmarked in terms of qubit relaxation times across the wafer. 32 sub-dies D1 and 24 sub-dies D2 from across a single 300 mm wafer were measured at 10 mK in a dilution refrigerator (without prior screening or pre-selection). A total of 394 out of 400 qubits (yield = 98.5%) were functional and fully characterized. The time averaged energy relaxation times ($\langle T_1 \rangle_t$) of the best performing qubits from sub-dies D1 show a center-to-edge dependence with $\langle T_1 \rangle_t^{\max}$ = 113 μs close the center and down to $\langle T_1 \rangle_t^{\min}$ = 42 μs near the edge with median of $\langle T_1 \rangle_t^{\text{med}}$ = 75 μs across the wafer (see wafer map in Figure 2a). Imposing radial symmetry, interpolated $\langle T_1 \rangle_t$ times as a function of distance from the center of the wafer are visualized with a background colormap in Figure 2a. The interpolation fits the data well, with only a few outliers deviating from the trend. This center-to-edge effect is commonly observed in most parameters controlled by wafer fabrication and indicates the importance of advanced process control. Since modern semiconductor tooling allows, in principle, such levels of control, our results also indicate a clear road forward towards even better, batch-to-batch identical devices in the future (Supplementary Figure 7). However, already on a per-die level (which matters most for quantum processors), the current results show a remarkable degree of performance.

To ensure objectivity and prevent cherry-picking, all reported $T_1$ values are time-averaged (on average 20 hours), with an example time trace depicted by Figure 2b. As commonly observed in transmon devices [33,34], we also report significant temporal variation ($\sigma \approx 20$ μs) in both relaxation and coherence times with near Gaussian distributions reaching $T_1$ = 161 μs and $T_2^e$ = 245 μs (inset in Figure 2b). Such temporal fluctuations are associated with coupling of the qubit to near-resonance two-level system (TLS) defects [35], which are in turn longitudinally coupled to low frequency (thermally active at 10 mK) two-level fluctuators (TLF) [33,36–38]. The example trace of Figure 2b, in addition, exhibits a jump in the average Hahn-echo coherence time $T_2^e$ of six standard deviations after 45 hours of measurements, without any known external trigger. The absence of a concurrent jump in $T_1$ indicates that the most plausible explanation is the vanishing of a coupled TLF (and not a near-resonant TLS), possibly due to a defect rearrangement from an impinging high-energy radiation event [39], or due to a trapped quasiparticle escaping from a shallow, local well in the superconducting gap [40].

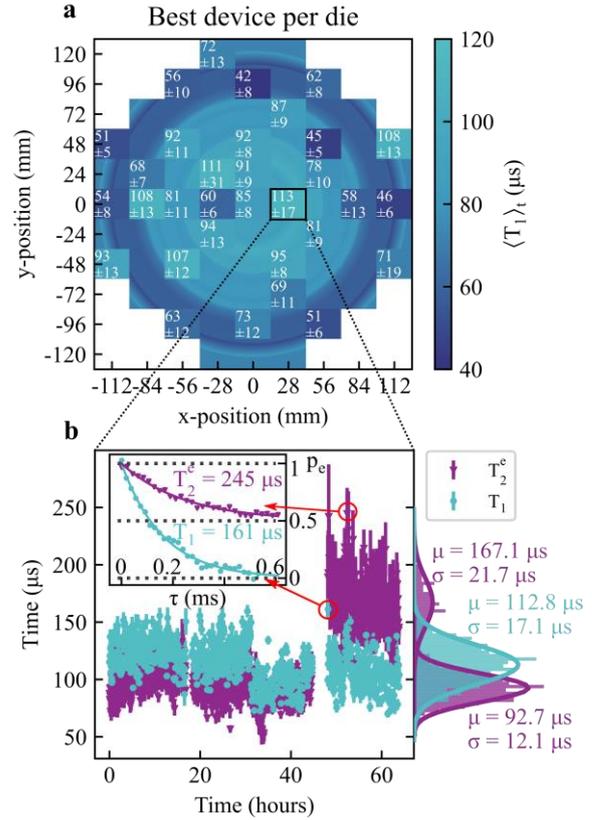

**FIG. 2 | Qubit relaxation and coherence times. a**, Wafer map of time averaged relaxation times $\langle T_1 \rangle_t$ of the best performing qubit on each measured die. The background color represents the interpolated average value as a function of radius. The mean values and standard deviations are printed at each measured location. **b**, Repeated measurements of the relaxation time $T_1$ and Hahn-echo coherence time $T_2^e$ of the best performing qubit on die (1,0) over a period of 60 hours. The time statistics are represented by histograms with Gaussian fits of mean μ and standard deviation σ. Inset: example traces of the qubit excited state population ($p_e$) relaxation and decoherence.

### IV. Coherence limitations and interface defects.

Following the comprehensive analysis of qubit quality across the wafer, a deeper exploration of qubit relaxation and coherence limitations is warranted. The transmon qubits on sub-dies D1 were designed with various capacitor geometry sizes, resulting in a six-fold difference in the extent of calculated electric-field-energy participation ratios (EPR) at qubit capacitor interfaces (metal-air, substrate-metal, and substrate-air) [41]. This allows for the differentiation between loss sources located at device interfaces and other losses (losses intrinsic to the JJ, bulk substrate loss, etc.). Qubit energy loss (1/Q) scales linearly with the sum of all interface EPR (Figure 3a), without signs of saturation, indicating that capacitor interface losses primarily govern qubit relaxation. A linear loss model fits the data well.

$$\frac{1}{Q} = \frac{1}{2\pi T_1 f_{qb}} = \delta_0 + \delta_{SA} p_{SA} + \delta_{SM} p_{SM} \quad (1)$$
$$+ \delta_{MA} p_{MA}$$
$$= \delta_0 + \delta_t (p_{SA} + p_{SM} + p_{MA})$$

With $\delta_0$ encompassing all qubit relaxation channels that are not located at device interfaces, $p_{SA}$ the substrate-air interface EPR, $p_{SM}$ the substrate-metal interface EPR, and $p_{MA}$ the metal-air interface EPR, all proportional to each other [41]. $\delta_{SA}, \delta_{SM}$, and $\delta_{MA}$ are the relaxation losses located at those respective interfaces, and $\delta_t$ represents the effective total interface loss. Projecting the total interface participation to zero yields a $T_1$ limit of $\sim 0.3^{\infty}_{0.15}$ ms at $f_{qb} = 3$ GHz. This is considerably greater than the mean values depicted in Figure 2a, affirming that, at present, qubit relaxation is predominantly dictated by capacitor interface losses. Furthermore, it implies that the overlap JJ fabrication, including the argon milling physical damage on the junction's BE, does not cause a noticeable rise in TLS defect losses within the Al/AlO$_x$/Al junction. This is consistent with previously observed resilience of Al to TLS defects induced by argon milling [32,42].

To gain insight into the coupled TLS defects, a magnetic flux tunable qubit on a sub-die D2 was used to scan the spectral environment. The qubit frequency as function of applied magnetic flux did not exhibit any observable avoided crossings with strongly coupled (> 250 kHz) TLS (data not shown), reminiscent of TLS located inside the JJ barrier [43,44]. However, individual, weakly coupled, TLS defects can be spectrally resolved with swap spectroscopy [38,45]. Here the qubit is excited with a π-pulse and then left to relax for a fixed duration while being detuned in frequency by a flux pulse, as shown by the pulse sequence of the inset of Figure 3b. This procedure is used with different flux pulse amplitudes to scan a frequency range of 500 MHz repeatedly for 13 hours.

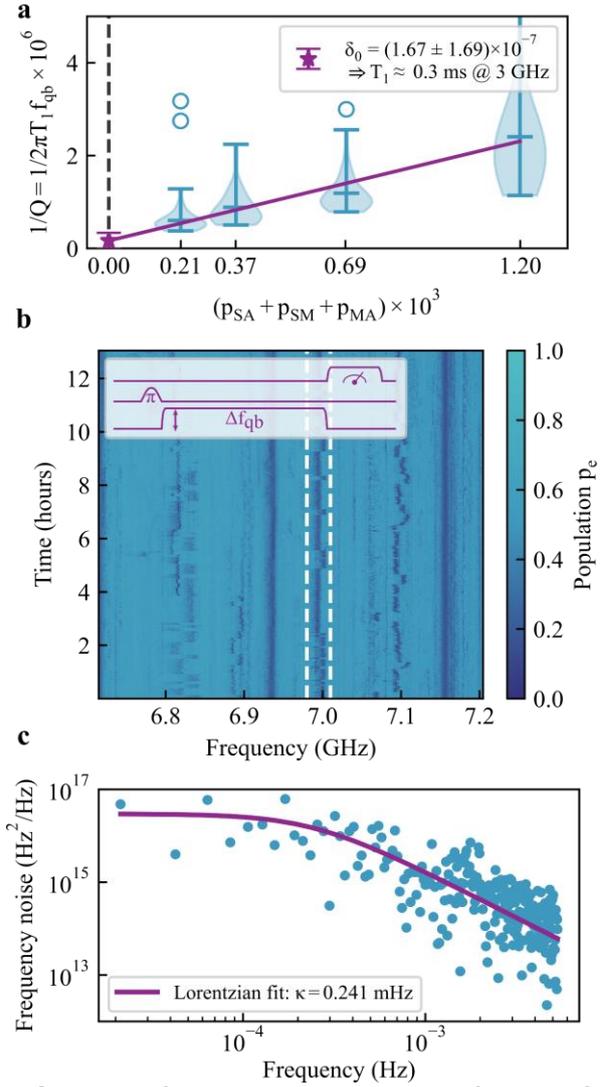

**FIG. 3 | Interface two-level system defects. a**, Qubit loss (1/Q-factor) as function of total interface (substrate-air, metal-air, and substrate-metal) participation ratio of electric-field-energy of four different qubit designs on sub-dies D1 across the wafer. The distributions of each design are filtered for outliers beyond the interquartile range and visualized with violin plots including the mean and extrema (outliers shown as scatter points). The linear fit extrapolates the loss to zero interface participation and extracts the upper bound on the relaxation time. **b**, A flux tunable qubit's population as a function of qubit frequency measured repeatedly over 13 hours to visualize the dynamics of relaxation channels associated with TLS defects at interfaces. Inset: the pulse sequence used to scan the spectrum. **c**, Frequency noise power spectral density of the highlighted defect in **b**.

The resulting time-frequency map of residual qubit population, visualized by Figure 3b, clearly shows several spectrally resolved relaxation channels. These relaxation channels are likely the weakly coupled TLS defects located at the capacitor interfaces, where the electric fields are orders of magnitude lower than inside the JJ [46,47]. Most of the defects show random-telegrapher-noise-like switching, attributed to coupling of these TLS with TLF [33,36–38]. We can estimate the density of TLS to be ~34/GHz (see Supplementary Figure 4) which is comparable to values reported for shadow evaporated qubits [46]. One of the fluctuating TLS was isolated and its frequency noise power spectral density (PSD) was calculated in Figure 3c. The frequency noise PSD of the TLS fits well with a single random TLF model characterized by a Lorentzian with TLF random switching rate of 0.241 mHz, comparable to timescales of previously reported relaxation time fluctuations [34]. The random jump in $T_2^e$ observed in Figure 2b could therefore be explained by the disappearance of similar TLF (with switching rate faster than the inverse measurement time) and consequential stabilization of the coupled TLS.

### V. Qubit frequency variability and aging.

The variability of qubit frequencies and JJ resistances are examined as indicators for the control and variability of the fabrication process. Transmon qubit frequency variability arises primarily from JJ critical-current variations.

The Ambegaokar-Baratoff relation [48] for the critical-current of a JJ describes the link between a transmon qubit's frequency [16] and the normal state resistance of the junction ($R_n$), $f_{qb} \propto \sqrt{1/R_n}$. Junction resistance is determined by the overlapping area ($A$) between the two junction electrodes, the barrier thickness ($d$) and barrier resistivity ($\rho$). Consequently, the qubit frequency variation is determined by the variations of these quantities. In the context of fabrication, JJ area variation is primarily controlled by patterning processes, while the uniformity of the barrier oxidation processing step regulates the variability of both the barrier resistivity and thickness. It is therefore prudent to monitor the variability of the JJ area, and the resistance-area product ($RA = \rho d$) which encapsulates resistivity and thickness into a single parameter, $RA$, representative of the overall oxidation uniformity (Supplementary Figure 5c).

Normal state resistance data of 7872 JJ test structures is collected across the wafer. Both the resistance values and the relative standard deviations (RSD) scale with estimated (Supplementary section II) junction area (Figure 4a,b). Approximately half of the variability can be attributed to a center-to-edge decrease of the average $R_n$ (see Figure 4c and Supplementary Figure 5d), based on the observed ~50% reduction in $\text{RSD}_{R_n}$ on a single die (pink data), compared to the across-wafer statistics (purple data) in Figure 4a. Furthermore, the JJ area dependence of the $R_n$ RSD fits well with a model of constant area variance ($\sigma_A^2$), derived by propagation of uncertainty [46,49].

$$\text{RSD}_{f_{qb}} = \frac{\text{RSD}_{R_n}}{2} \approx \frac{1}{2}\sqrt{\text{RSD}_{RA}^2 + \frac{\sigma_A^2}{A^2}} \quad (2)$$

Fitting the data to equation (2) allows for the disentanglement of barrier non-uniformity and area variability. This analysis reveals an $\text{RSD}_{RA}$ = 4.47% and $\sigma_A = 0.00334\ \mu m^2$ on a single die, meaning that, for all JJ with $A > 0.075\ \mu m^2$, the barrier non-uniformity is the dominant cause of $f_{qb}$ variability. The area standard deviations are comparable with values reported in literature [46,50](see Supplementary section VIII for more details).

The relative standard deviation (RSD) of the measured qubit frequencies on sub-dies D2 across the wafer (excluding the magnetic flux tunable qubit) compares well with the expected relation $RSD_{R_n} = 2RSD_{f_{qb}}$ (following $f_{qb} \propto \sqrt{1/R_n}$). The $RSD_{f_{qb}}$ ranges between 5% and 7% for the different junction areas tested (Figure 4a), with no discernible trend with area. Across the wafer, locally averaged qubit frequencies (including JJ area correction $f_{qb}/\sqrt{A} \propto 1/\sqrt{RA}$) exhibit a similar center-to-edge increase (Figure 4c), also in agreement with JJ test array $R_n$ wafer maps (Supplementary Figure 5d).

Leveraging the possibility of even more advanced process control, as present in modern day tooling, we can envisage ways to further push the qubit frequency homogeneity towards across-wafer, die-to-die consistent values. Our analysis indicates that significant variability is currently coming from the controlled oxidation process, which can still be further fine-tuned. Residual area variance can be addressed with more advanced lithographic control tricks, and etch fine tuning, in principle beyond the range currently achievable with angled shadow evaporation techniques [46,51]. Furthermore, higher order contributions to the JJ resistance variation e.g. aluminum grain-size effects, and sidewall-edge-slope (impacted by etch and argon milling) could be further addressed. Finally, at design level, larger junction areas (with higher $RA$ values) can be targeted, minimizing junction size variation, consistent with recent literature reports [46].

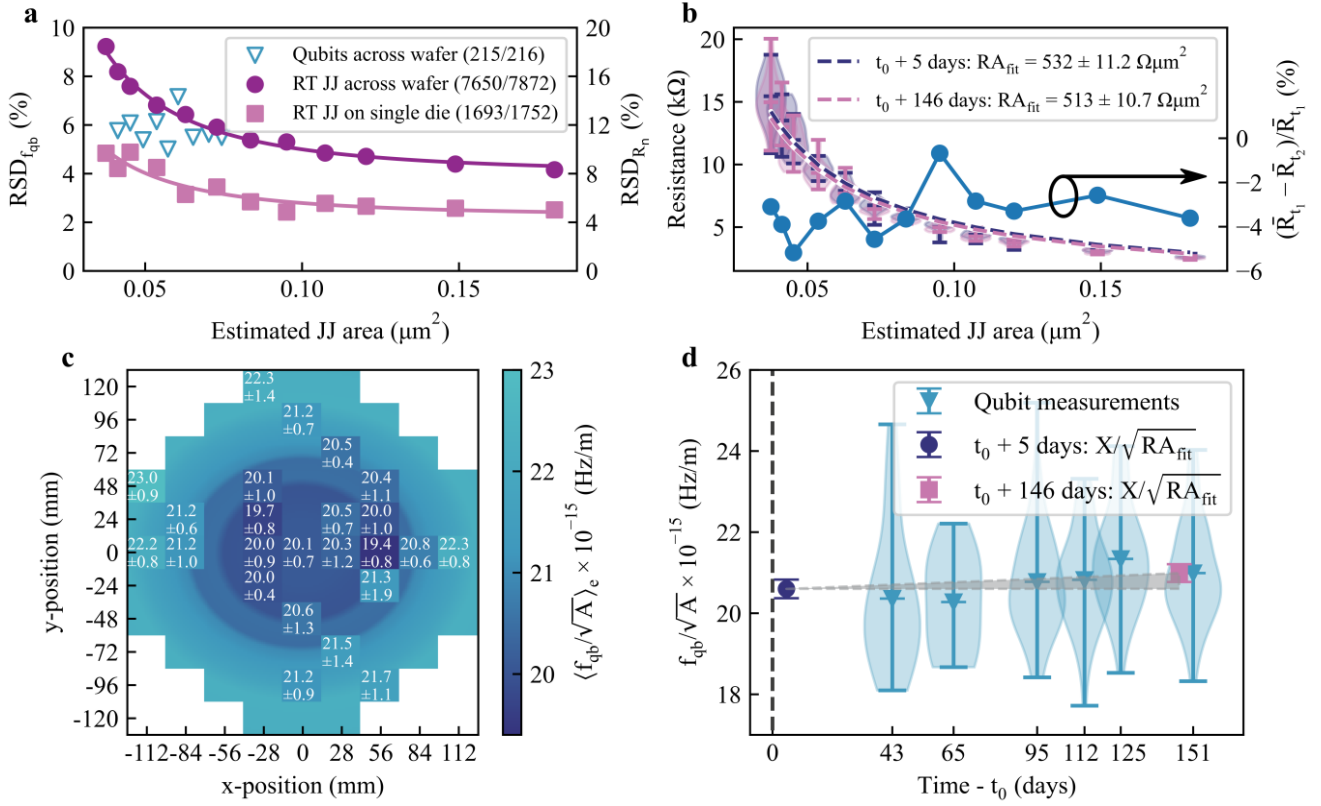

**FIG. 4 | Qubit frequency variability and aging analysis. a**, Relative standard deviation of qubit frequencies and normal state resistances of JJ test structures ($\text{RSD}_{f_{qb}} = \text{RSD}_{R_n}/2$, visualized with double y-axis) as a function of the estimated JJ area. 9 (qubits/die) × 24 (dies) = 216 qubits across wafer, 8 (JJ/area/die) × 12 (areas) × 82 (dies) = 7872 JJ across wafer, 146 (JJ/area) × 12 (areas) = 1752 JJ on die (0,2). All working qubits are included, JJ test structures resistances were filtered per area for outliers beyond 1.5 times the interquartile range. Solid lines are fits with equation (2). **b**, JJ test structure normal resistances as a function of estimated junction area for a subgroup of 216 devices measured 5 days after fabrication ($t_0$) and once more 146 days after fabrication. The right y-axis shows the corresponding relative average resistance change for each JJ area. **c**, Wafer map of qubit frequencies scaled with $\sqrt{A}$ ($A$ the estimated JJ area) and ensemble averaged over the nine qubits of different JJ area on each sub-die D2. The mean values are printed, accompanied by the standard deviation. The background color represents the interpolated average value as a function of radius. **d**, The area scaled qubit frequencies of qubits on sub-dies D2 plotted as function of cooldown time (distributions of measured qubits represented by violin plots with means and extrema), compared with the JJ resistance area product values extracted from the fits in **b** scaled with a proportionality factor $X \approx \sqrt{\Delta E_C/he^2}$ (with $\Delta$ the superconducting gap of aluminum, $E_C$ the qubit charging energy, $h$ Planck's constant, and $e$ the elementary charge). The black vertical dashed line represents the wafer fabrication date at $t_0$.

Shadow evaporated JJ often show notable parameter aging (increase in the normal resistance as a function of time) with reported values as high as tens of percents, over periods spanning days to weeks [46,52,53]. Overlap junctions fabricated in this work, however, show significantly less aging, limited to a decrease of 3.7% over 146 days (Figure 4b). A subset of 216 JJ test structures was measured five days post-fabrication, and once more 146 days following fabrication (the diced wafer was stored in the cleanroom environment at RT, covered by protective resist). Furthermore, junction aging can also be tracked by comparing $f_{qb}/\sqrt{A}$ ($\propto 1/\sqrt{RA}$) of sub-dies D2 as a function of their cooldown date in the span of 151 days (Figure 4d). The aged $RA$ values, obtained from the JJ test structure measurements (rescaled to match the qubit values) illustrate that any detected aging falls within the measured qubit frequency variation. Aging can be linked to the presence of barrier impurities [54], or oxygen diffusion [46,55]. The presented overlap JJ fabrication process has no organic material present during the formation of the barrier, in contrast to the conventional fabrication process where organic photoresist is used as the angled shadow mask. We hypothesize therefore that the weak aging effects in the JJ resistance of the presented process could potentially be attributed to the absence of carbon impurities within the barrier (see Supplementary Figure 6), though further studies would be required to substantiate that hypothesis.

## VI. Conclusion and outlook.

In summary, our study presents for the first time a large-scale fabrication process for superconducting qubits using fully industrial semiconductor nanofabrication methods on 300 mm silicon wafers, achieving high coherence and across-wafer yield of 98.5% in an industry-standard facility. The process quality was confirmed through large scale statistics of qubit relaxation and coherence measurements conducted across the wafer, including time-averaged $T_1$- and $T_2^e$-times exceeding 100 $\mu$s. Our measurements shed light on observed center-to-edge dependencies suggesting an avenue for further improvements leveraging the advanced process control of modern semiconductor nanofabrication tooling. Our investigation identified the dominant relaxation and decoherence sources as TLS defects located at the device interfaces. These two observations allow for further in-depth optimizations that should result in additional improvements in qubit quality. In addition, we characterized the variability of the process using JJ normal resistance measurements and assessed qubit frequency variability. Our results indicate that the current limitation resides in barrier oxidation. However, the area control realized through optical lithography and subtractive dry etching shows clear promise for continued optimizations beyond the state-of-the-art. Finally, we have verified the stability of our process over a period of at least 151 days. This underscores the robustness of the fabrication process and its suitability for reliable large-scale quantum processor fabrication.

## Acknowledgements

The authors are grateful for the support of the imec P-line, operational support, the MCA team, Koen Verhemeldonck for wire-bonding support, and Fred Loosen for taking pictures. Arik Mahbub is acknowledged for supporting JJ resistance analysis. This work was supported in part by the imec Industrial Affiliation Program on Quantum Computing. J.V.D. acknowledges the support of the Research Foundation-Flanders (FWO) through the SB Ph.D. program (Grant No. 1S15722N). The authors would also like to thank O. Painter, J. Bylander, C. Haffner, and P. McMahon for insightful comments on this work.

## Author's contribution

S.M. coordinated the fabrication process, with etch development by Y.C., thin film development by D.P.L., lithography development by D.V. and Y.H., and clean development by J.G.L.. Ts.I. developed the post-processing and sample preparation procedure. A.P., M.M., and J.V.D. designed the qubit samples. J.V.D., R.A., M.D. and A.P. performed the measurements and analysis of qubit data at cryogenic temperatures. S.M., and Ts.I. performed the measurements of the JJ test arrays at room temperature. J.V.D. analyzed the gathered JJ test array data. J.V.D., R.A., A.P., M.M., and A.M.V. prepared the experimental setup and methods. J.V.D. prepared the manuscript, with input from all authors. A.P., M.M., D.W., J.D.B., and K.D.G. supervised and coordinated the project.

## Data availability

The data that support the figures and conclusion within this paper and other finding of this study are available upon reasonable request.

# Supplemental material

## I. Device designs

The mask design of one die on the wafer in this work constitutes 20 different sub-dies of mostly fixed frequency qubits coupled to read-out resonators and a common feedline. A large parameter space is covered with these designs to accommodate the a priori unknown values of co-planar waveguide (CPW) phase velocities, JJ barrier resistivity and thickness of the new process. Two sub-die designs D1 and D2 were selected for this study, with their optical microscope pictures shown in Figure S1 and their design parameters summarized by Table SI.

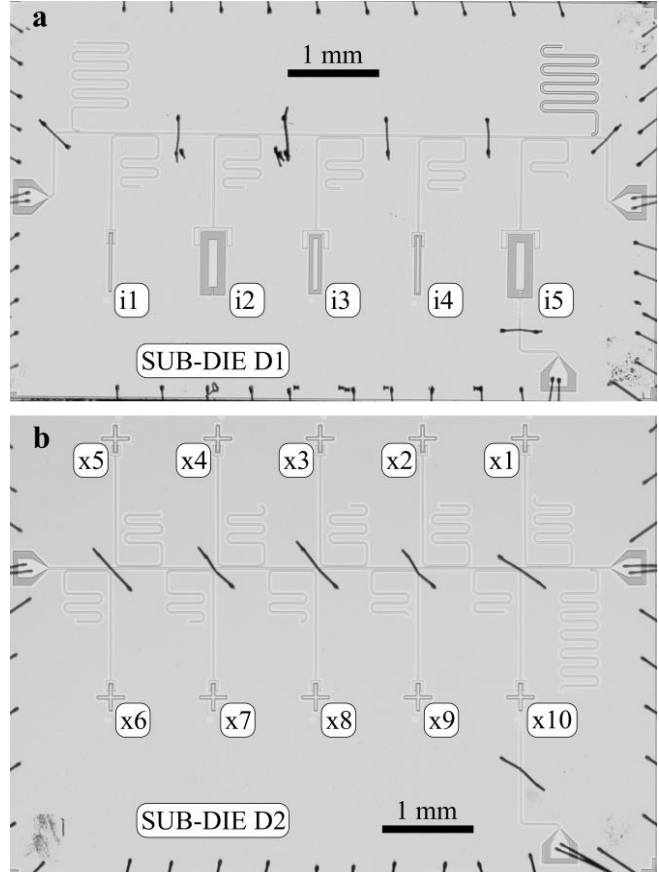

SUPPLEMENTARY FIG. 1. **a**, Optical micrograph of an example sub-die D1, wire-bonded into an aluminum measurement package. **b**, Optical micrograph of an example sub-die D2, wire-bonded into an aluminum measurement package.

TABLE I. **Qubit design parameters.** The qubits on sub-dies D1 and D2 as labeled in Supplementary Figure 1 were designed with the parameters summarized in this table. The planar capacitor width and gap to the ground plane, the critical junction dimension (cd), the CPW read-out resonator frequency ($f_r$), the qubit frequency ($f_{qb}$), the charging energy of the transmon ($E_C$), the Josephson energy of the transmon ($E_J$), the coupling strength between qubit and read-out resonator (g) and if a symmetric superconducting quantum interference device (SQUID) is used to make the qubit flux tunable.

| Name | width (μm) | gap (μm) | cd (nm) | $f_r$ (GHz) | $f_{qb}$ (GHz) | $\frac{E_C}{h}$ (MHz) | $\frac{E_J}{h}$ (GHz) | g (MHz) | SQUID |
|---|---|---|---|---|---|---|---|---|---|
| i1 | 13 | 13 | 120 | 7.2 | 3.94 | 203 | 10.56 | 77.7 | no |
| i2 | 90 | 90 | 90 | 6.3 | 2.90 | 202 | 5.94 | 55.0 | no |
| i3 | 48 | 48 | 100 | 6.6 | 3.27 | 206 | 7.33 | 68.8 | no |
| i4 | 24 | 24 | 110 | 6.9 | 3.59 | 203 | 8.87 | 68.0 | no |
| i5 | 90 | 90 | 110 | 7.5 | 5.15 | 202 | 17.75 | 85.6 | yes |
| x1 | 20 | 13 | 119 | 6.0 | 3.08 | 231 | 5.94 | 52.5 | no |
| x2 | 20 | 13 | 131 | 6.2 | 3.45 | 231 | 7.33 | 57.1 | no |
| x3 | 20 | 13 | 143 | 6.4 | 3.82 | 231 | 8.87 | 61.8 | no |
| x4 | 20 | 13 | 156 | 6.6 | 4.19 | 231 | 10.56 | 66.5 | no |
| x5 | 20 | 13 | 168 | 6.8 | 4.48 | 231 | 12.02 | 70.7 | no |
| x6 | 20 | 13 | 180 | 7.0 | 4.74 | 231 | 13.37 | 74.7 | no |
| x7 | 20 | 13 | 191 | 7.2 | 5.11 | 231 | 15.42 | 79.6 | no |
| x8 | 20 | 13 | 205 | 7.4 | 5.47 | 231 | 17.62 | 84.5 | no |
| x9 | 20 | 13 | 220 | 7.6 | 5.77 | 231 | 19.48 | 88.9 | no |
| x10 | 20 | 13 | 165 | 7.8 | 6.07 | 231 | 21.47 | 93.4 | yes |

## II. JJ area estimation

The actual JJ areas deviate significantly from the designed values. A challenging aspect of the dimension targeting arises from the tapering of the BE sidewalls (followed by argon milling), necessary for good step coverage of the TE deposition. The process flow presented in this work can be further optimized with, for example, etch process fine tuning and optical proximity corrections for improved critical dimension (cd) targeting. The JJ analysis delineated in the main text utilizes a "best-effort" area estimate for the JJ. First an offset on the designed cd (the BE width and TE width of the junction) is determined from SEM images taken across the wafer, with examples illustrated by Supplementary Figure 2a,b. In Supplementary Figure 2d, a constant cd correction for BE and TE is extracted from a fit to the SEM inspections. Secondly, the rounded surface of the BE due to the argon milling process is accounted for with an approximate formula for the circumference of an ellipse. The major radius of the ellipse is gathered from the BE cd, while the minor radius is extracted from the TEM image in Supplementary Figure 2c ($h_{BE} = 59$ nm). Supplementary Figure 2e illustrates the "best-effort" area estimate for each designed nominal cd. The "best-effort" approach allowed us to fit the JJ normal resistance as function of area (main text Figure 4b) with only one free fitting parameter "RA", and no need for any further area corrections or offset series resistance (besides the included resistance of 32 $\Omega$ measured on shorted test structures to account for probe contact resistance). Note that any location dependence of the area, although observed (Supplementary Figure 5), is not included for the area estimates in the main text, due to insufficient data for all locations. The observed center-to-edge drift is discussed in more detail below.

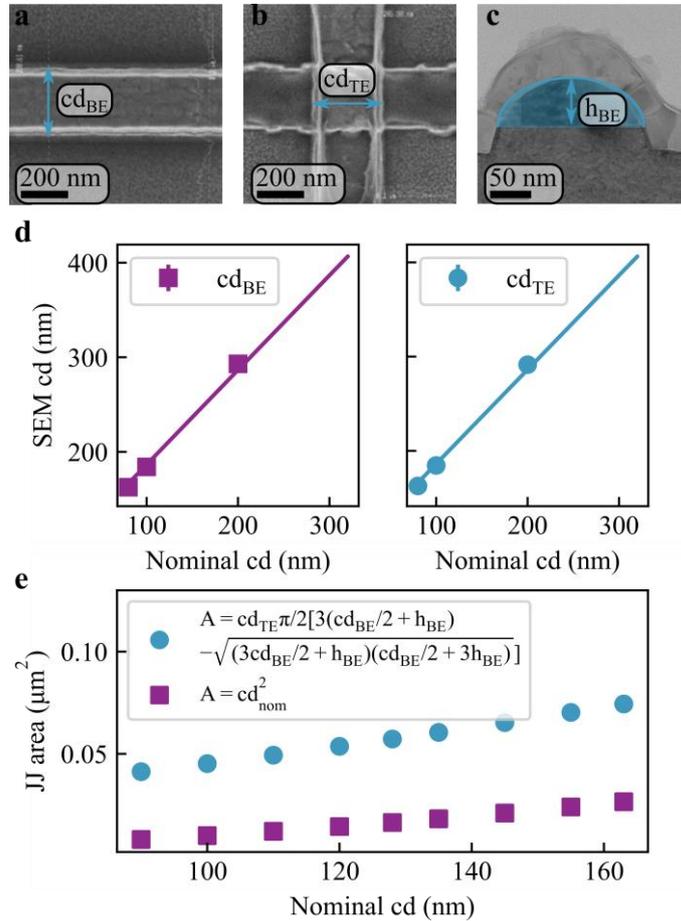

SUPPLEMENTARY FIG. 2. **JJ area estimation. a**, SEM image of a JJ after BE patterning. **b**, SEM image of a JJ after TE patterning. **c**, TEM image of a cross section of a JJ, including an ellipse circumference as approximation of the rounded BE surface. **d**, Average width of the JJ's BE and TE measured from the SEM images a,b. The solid lines are linear fits including a constant offset. **e**, Best effort estimate of the JJ area, calculated for each designed JJ critical dimension.

## III. Interface participation ratio

The different qubit designs on sub-die D1 have planar capacitors resembling a section of CPW (Supplementary Figure 1a) allowing for a simplified analytical calculation of the electric field energy participation at the interfaces [41] of the qubit capacitor. The substrate-air (SA), metal-air (MA), and substrate-metal (SM) interface participation ratios are calculated for the four different capacitor geometries (width and gap ∈ $\{13, 24, 48, 90\}$ μm). We used interface thicknesses of 5 nm for the MA and SM interfaces, and 3 nm for the SA interface. The relative dielectric constants used in the calculations are $\epsilon_{sub} = 11.9$ (for the Si substrate), $\epsilon_{MA} = 10$ (for the Al oxide), $\epsilon_{SM} = 11.9$ (for the Si-Al interface), and $\epsilon_{SA} = 3.9$ (for the SiO$_2$). The calculated interface participation ratios for the four different qubit geometries are plotted in Supplementary Figure 3.

The calculated participations of the SA, MA, and SM interfaces are proportional to each other, which allows us to group them into a total interface participation of our simple loss model of equation (3) in the main text. Note that the performed analysis and drawn conclusions do not rely on the exact values of the calculated participation ratios, only on the total interface contribution dependence on the different device geometries.

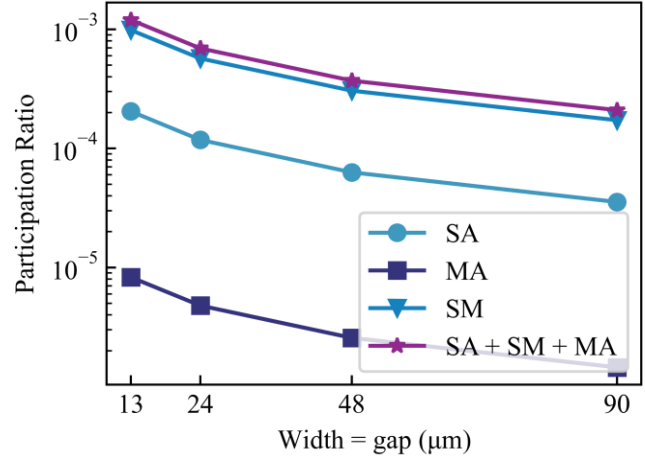

SUPPLEMENTARY FIG. 3. The calculated electric field energy participation ratios calculated for the four different qubit capacitor geometries on sub-die D1.

## IV. Two-level system density

An estimated two-level system (TLS) density is extracted from the time-frequency map of main text Figure 3b. The data is first sliced along the time-axis into frequency traces, exemplified in Supplementary Figure 4. The number of TLS in each slice is counted using the python "scipy.signal.find_peaks()" method. The total number of counted TLS is then averaged over all timestamps, resulting in a density of 34.4 TLS/GHz. We note that the number of identified TLS, and therefore the extracted density, depends on the settings of the peak finding algorithm and was chosen such that the highlighted map of Supplementary Figure 4a looked satisfactory.

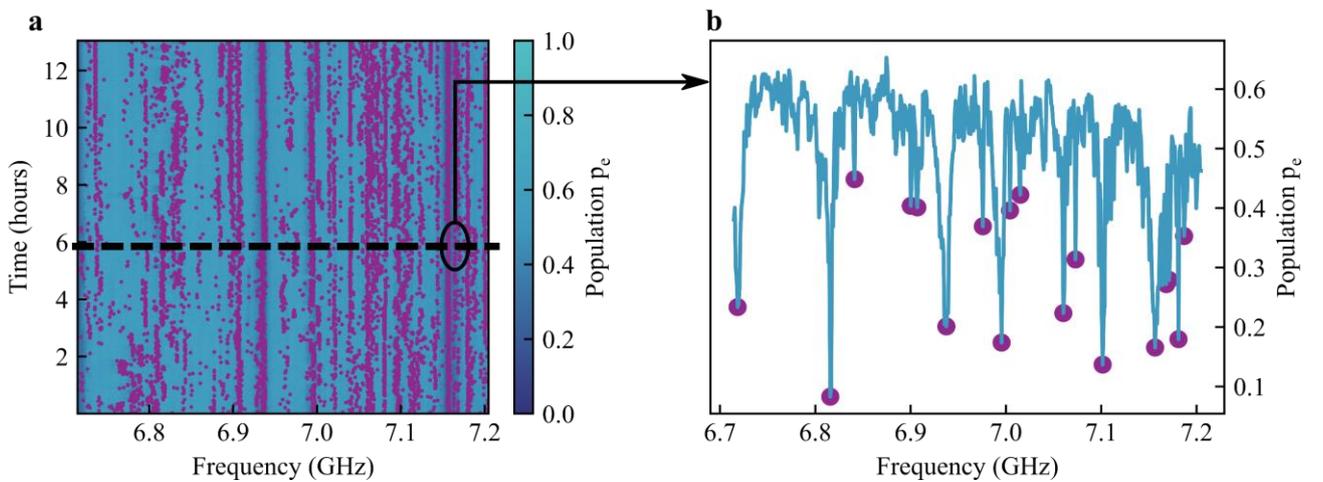

SUPPLEMENTARY FIG. 4 **Two-level system counting. a**, Time-frequency map (main text Figure 3b) with highlighted two-level systems detected using a python peak search algorithm (method called: "scipy.signal.find_peaks(prominence=0.15)"). **b**, Example slice of **a** illustrating the detected TLS at one timestamp.

# V. Josephson junction normal resistance wafer location dependence

A clear center-to-edge variation is observed in the qubit frequencies measured across the wafer (main text Figure 4c). This trend is shared with the normal resistance measurements performed on Josephson junction (JJ) test arrays across the wafer (Supplementary Figure 5d). Sampled inspections of JJ electrode dimensions across the wafer (Supplementary Figure 5a,b) reveal that some, but likely not all (more data and statistics are required), of the location dependence results from etch non-uniformity between the center and the edge of the wafer. The barrier oxidation process, on the other hand, does not show such a center-to-edge dependence when replicated on un-patterned, blanket wafers, as illustrated by the resistance-area product wafer-map in Supplementary Figure 5c. This analysis clearly identifies the patterning process as a contributor to the center-to-edge dependence of qubit frequencies measured across the wafer. However, other sources like bottom electrode morphology and roughness (determined by etch and argon milling), or aluminum grain size are likely to exhibit center-to-edge variation as well. Large-scale inspections of device critical dimensions are now being developed to quantify the contributions of each processing step to the center-to-edge variation and improve the fabrication methods of this work towards higher wafer-uniformity.

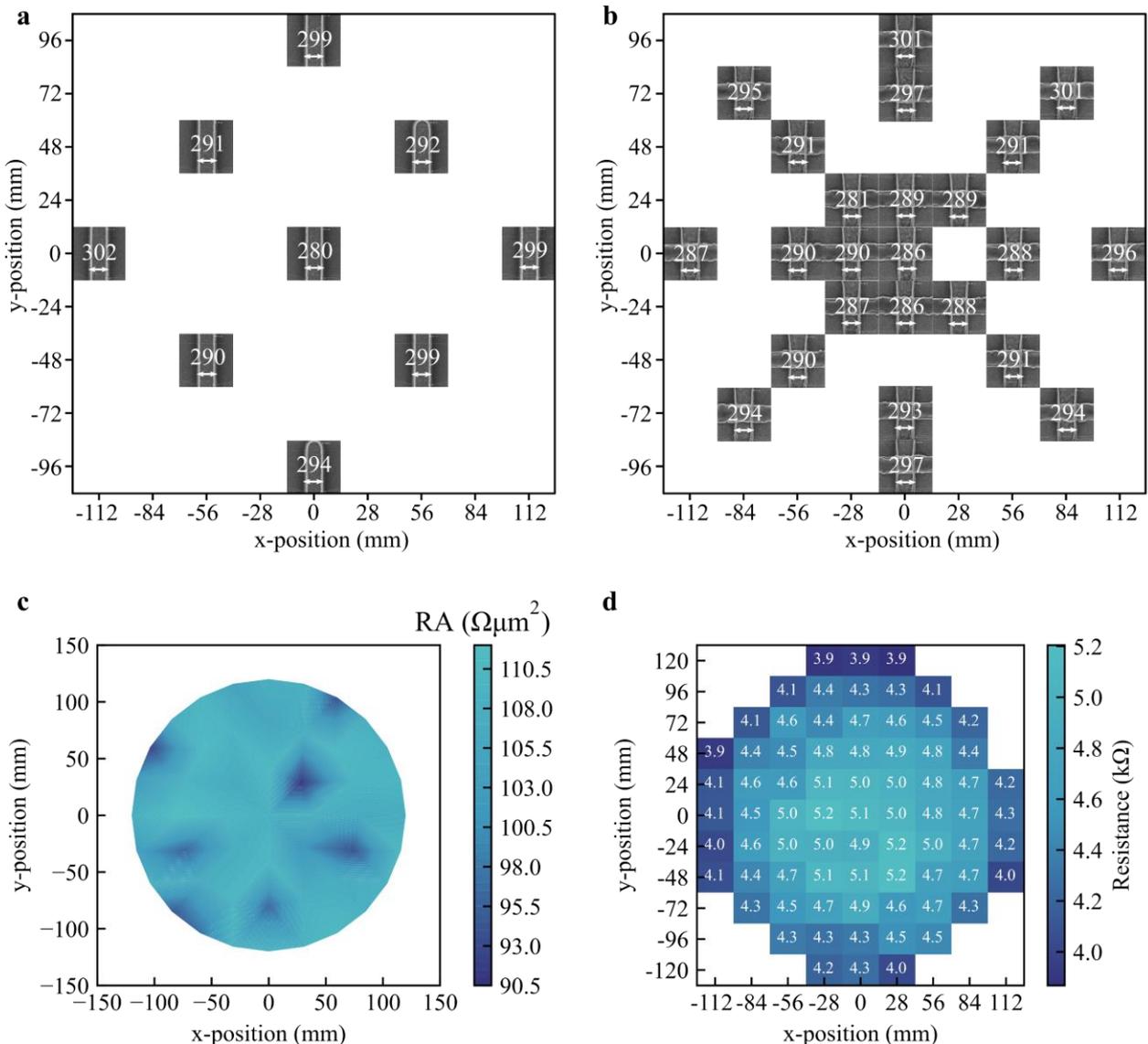

SUPPLEMENTARY FIG. 5 **JJ resistance variability wafer location dependence. a**, Scanning electron microscope images of the bottom electrode of junctions across the wafer (designed width = 200 nm). The measured width of the electrode (in nm) is annotated. **b**, Scanning electron microscope images of the top electrode of junctions across the wafer (designed width = 200 nm). The measured width of the top electrode (in nm) is annotated. **c**, Resistance-area product extracted via current in-plane tunneling (CIPT) characterization on a wafer where the barrier oxidation process was replicated on blanket without patterning. **d**, Average normal resistance of Josephson junctions (8 copies per die), with designed electrode widths of 200 nm, measured across the wafer.

The reported scanning electron microscope (SEM) images in Supplementary Figure 5a,b were taken on the 300 mm wafer using a Hitachi CD-SEM tool with included dimension measurements. The SEM image of main text Figure 1c was taken with a HELIOS 1200AT tool. The Resistance-area product on the blanket wafer (Supplementary Figure 5c) were extracted using current in-plane tunneling (CIPT) on a Capres microHall-A300 (MH300) with probe L8pp.A03. The JJ test array normal resistances were measured at room temperature with a Tokyo Electron SYSTEM VI PRECIO tool. Additional JJ test-array normal resistance measurements were performed on diced coupons with a SUSS MicroTec iVista high-resolution wafer prober (data used for the single-die statistics of main text Figure 4a and aged data of Figure 4b,d).

## VI. Josephson junction energy dispersive X-ray analysis

The composition of the overlap Josephson junctions (JJ), fabricated with the process described in this work, is investigated with energy dispersive X-ray (EDS) analysis. A focused ion beam (FIB) specimen is extracted with a Helios 450, HP FIB lift-out system, and further analyzed with a scanning transmission electron microscope (STEM) Titan G2 at 200 kV. Like in our previous work [26] traces of Si and Ar are detected inside the junction barrier, while no C is present.

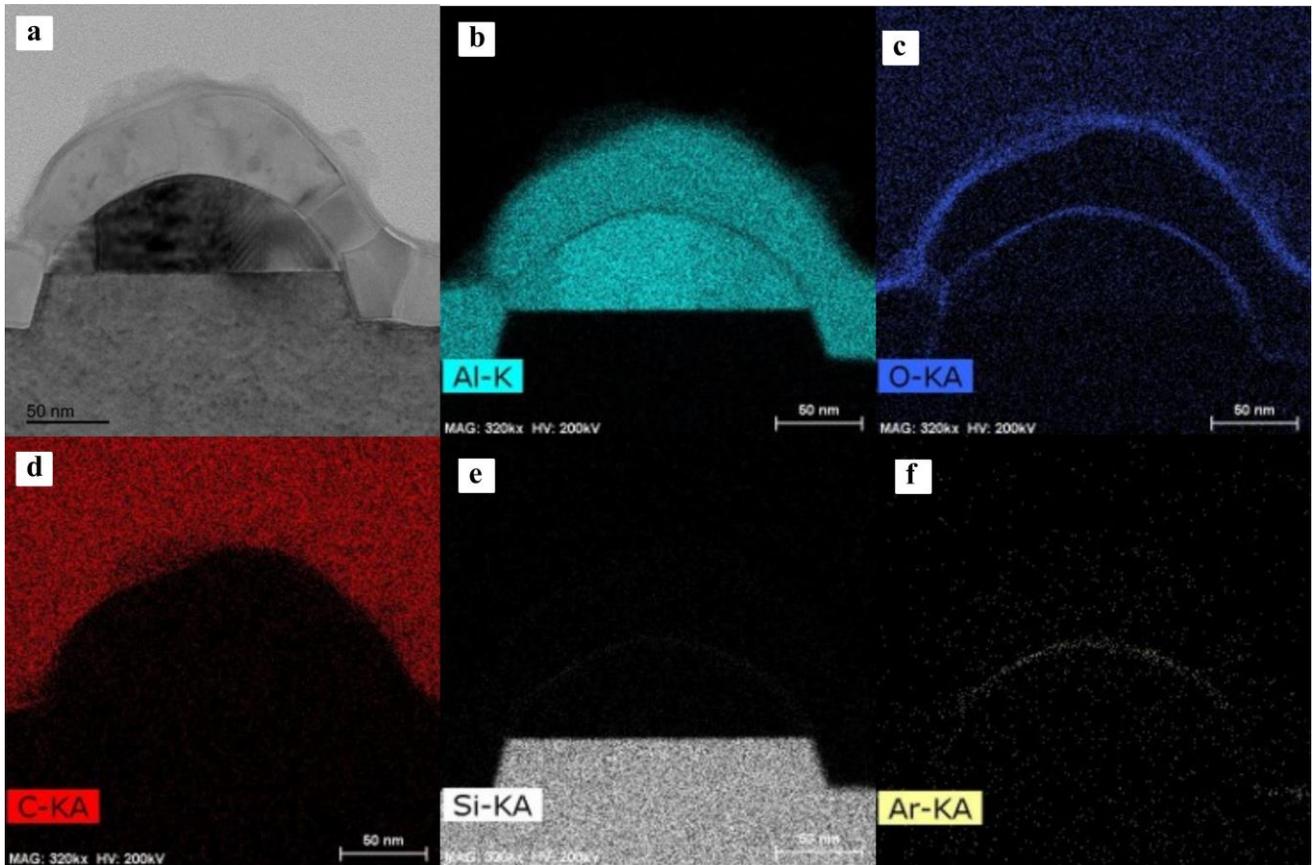

SUPPLEMENTARY FIG. 6. **a**, Scanning transmission electron microscopy (STEM) image of the cross-section of a JJ. **b,c,d,e,f** Energy dispersive X-ray (EDS) analysis of the chemical composition of the JJ cross-section. The detected atomic concentration of Al, O, C, Si, and Ar is colored respectively.

## VII. Batch-to-batch reproducibility

The data presented in the main text is collected from one single wafer (Batch 1, Wafer 1). In Supplementary Figure 7 we demonstrate with JJ normal resistance measurements that the described fabrication process can be reproduced reasonably well, both in terms of batch-to-batch and within-batch comparisons. We define a batch as a set of wafers that are processed sequentially at each step of the process, without major time interruption or, with identical tool settings. We note that currently our efforts prioritize process optimization in terms of qubit metrics, while more extensive wafer-to-wafer and batch-to-batch reproducibility studies, including cryogenic qubit measurements, are planned for future works.

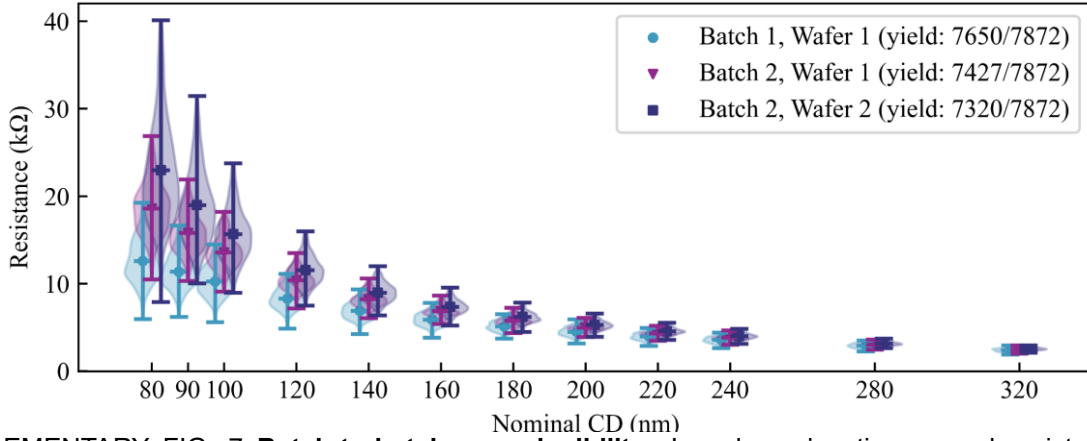

SUPPLEMENTARY FIG. 7 **Batch-to-batch reproducibility.** Josephson junction normal resistance values measured on 7872 test structures across each wafer. The distribution of resistances at each designed junction lead critical dimension (CD) is filtered for outliers beyond 1.5 times the interquartile range. The filtered distributions are visualized with violin plots including the mean and extrema at each CD (offset included for improved visibility).

## VIII. JJ normal resistance relative standard deviation analysis

The relative standard deviation of the JJ normal resistance shows a functional dependence on the JJ area, as shown in Figure 4a of the main text. Modeling this area dependence can be leveraged to disentangle variance contributions of area and barrier uniformity (tracked with the resistance area product parameter $RA = \rho d$). Propagation of uncertainty is used on $R_n = RA/A$ to arrive at $RSD_{R_n} \approx \sqrt{RSD_{RA}^2 + RSD_A^2}$, for which a logical assumption to make is that the barrier non-uniformity is independent of JJ area. The observed area dependence of $RSD_{R_n}$ is then entirely attributed to $RSD_A = f(A)$. The area of the JJ fabricated in this work includes the rounded surface of the bottom electrode and is described by the formula in the legend of Supplementary Figure 2e. The total area variance $\sigma_A$ of these junctions have contributions from variations in the bottom electrode critical dimension ($CD_{BE}$), variations in the top electrode dimension ($CD_{TE}$), and variations of the bottom electrode thickness ($h_{BE}$). At present, we do not have sufficient data to model the area variance in its entire functional dependence $\sigma_A = f'(CD_{BE}, CD_{TE}, h_{BE}, A)$, so we consider two models. In model A, we assume that the area variance is area independent $d\sigma_A/dA = 0$, such that $RSD_A = \sigma_A/A$. In model B, we oversimplify the area calculation as $A = (CD)^2$, with $CD = CD_{BE} = CD_{TE}$ and $d\sigma_{CD}/dA = 0$, such that $RSD_A = 2\sigma_{CD}/\sqrt{A}$, similar to literature [46,49].

In Supplementary Figure 8a we fitted both models to the measured JJ normal resistance RSD data (across wafer and on one single die) of Figure 4a in the main text, with best fit parameter values reported in Table II. Both models result in fitted curves that represent the data well, however the extracted fit uncertainties of model A are considerably lower than those of model B (see Table II). Therefore, we proceeded, in the main text, with our analysis using model A. One important remark to make here is that, according to model A, the contribution of the barrier non-uniformity to the measured resistance variability is dominant already for JJ with $A > 0.075$ $\mu$m$^2$, while according to model B, all measured structures are currently dominated by area variability. This is a conclusion drawn from Supplementary Figure 8b, where we calculate and compare the $RSD_A$ for both models on the data collected on a single die.

TABLE II. Best fit parameter values of models A and B fitted to the data in Supplementary Figure 8. The values are reported together with one standard deviation in fit uncertainty.

|         |              | RSD$_{RA}$ (%)            | $\sigma_A$ ($\mu$m$^2$)          | $\sigma_{CD}$ (nm) |
|---------|--------------|---------------------------|----------------------------------|--------------------|
| Model A | across wafer | 7.94 $\pm$ 0.18           | 0.00608 $\pm$ 9.9 $\times$ 10$^{-5}$ |                    |
|         | single die   | 4.47 $\pm$ 0.37           | 0.00334 $\pm$ 2.0 $\times$ 10$^{-4}$ |                    |
| Model B | across wafer | 5.87 $\times$ 10$^{-6}$ $\pm$ 6944 |                          | 16.5 $\pm$ 0.51    |
|         | single die   | 0.895 $\pm$ 3.2           |                                  | 9.07 $\pm$ 0.62    |

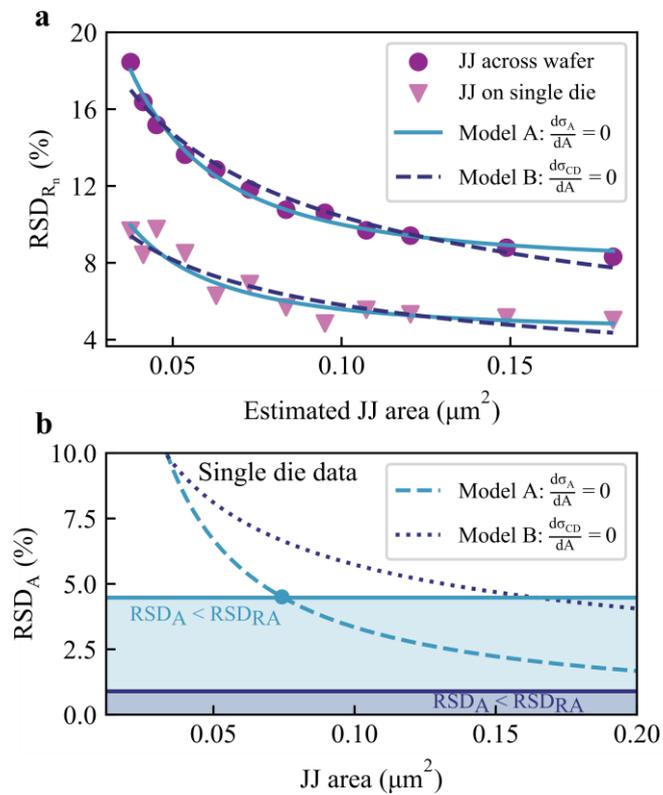

SUPPLEMENTARY FIG. 8 **Resistance variance area dependence model comparison. a**, Relative standard deviation of JJ normal resistances measured across the wafer and on a single die as function of the estimated JJ area (same data as Figure 4a). The solid lines represent the best fit with model A of constant area variance ($d\sigma_A/dA = 0$), the dashed line is the best fit with model B of constant critical dimension variance ($d\sigma_{CD}/dA = 0$). The corresponding fit parameter values are presented in Table II. **b**, The relative standard deviation of the JJ area as function of area is calculated for both models from the best fit values in Table II (only single die data shown for figure visibility). The horizontal lines highlight the cross-over from area variance dominated ($RSD_A > RSD_{RA}$) to barrier uniformity limited ($RSD_{RA} > RSD_A$) RSD of the JJ normal resistance for model A and model B.